\documentclass{osa-article}

\usepackage{amsfonts}
\usepackage{amsmath}

\usepackage{amssymb}
\usepackage{booktabs}
\usepackage{tabularx}
\usepackage{mathtools}
\usepackage{pgfplots}
\usepackage{geometry}
\usepackage{pdflscape}
\usepackage{enumerate}
\usepackage{xcolor}
\usepackage{esvect}
\usepackage{afterpage}
\usepackage[figuresright]{rotating}
\usepackage{footnote}
\usepackage{listings}
\usepackage{xcolor}
\usepackage{enumitem}
\usepackage{tocloft}
\usepackage{textcomp}
\usepackage{braket}
\usepackage{nicefrac}
\usepackage{nccmath}
\usepackage{bm}
\usepackage{import}
\usepackage{dpfloat}
\usepackage{nicefrac}
\usepackage{subcaption}
\usepackage{float}
\usepackage{afterpage}
\usepackage{booktabs} 
\usepackage{multirow}
\usepackage{siunitx}
\usepackage{graphicx}
\usepackage{tikz}
\definecolor{uniRed}{RGB}{171,31,45}
\definecolor{uniStone}{RGB}{190,185,166}
\definecolor{graphColor1}{HTML}{42145f}
\definecolor{graphColor2}{HTML}{002f5f}
\definecolor{graphColor3}{HTML}{003d4c}
\definecolor{graphColor4}{HTML}{024731}
\definecolor{graphColor5}{HTML}{53682b}
\definecolor{graphColor6}{HTML}{86431e}
\definecolor{graphColor7}{HTML}{5e3032}
\definecolor{graphColor8}{HTML}{772059}
\pgfplotsset{compat=newest}
\usetikzlibrary{pgfplots.colorbrewer}
\usetikzlibrary{plotmarks}
\usetikzlibrary{arrows}
\usetikzlibrary{spy}
\usetikzlibrary{shapes}
\usetikzlibrary{positioning}
\usetikzlibrary{calc}
\usetikzlibrary{decorations.pathreplacing}
\usetikzlibrary{decorations.pathmorphing}
\usetikzlibrary{shapes.geometric}
\usetikzlibrary{shapes}
\usetikzlibrary{shapes.symbols}
\usetikzlibrary{arrows.meta}
\makesavenoteenv{tabular}

\DeclarePairedDelimiter\abs{\lvert}{\rvert}%
\DeclarePairedDelimiter\norm{\lVert}{\rVert}%
\makeatletter
\let\oldabs\abs
\def\abs{\@ifstar{\oldabs}{\oldabs*}}
\let\oldnorm\norm
\def\norm{\@ifstar{\oldnorm}{\oldnorm*}}
\makeatother

\usepackage[ruled,resetcount]{algorithm2e}
\makeatletter
\renewcommand*\l@algocf{\l@figure}
\makeatother
\SetAlgoInsideSkip{smallskip}

\definecolor{light-gray}{gray}{0.95}

\usepackage{graphicx}
\newcommand{\CC}{%
    {\settoheight{\dimen0}{C}C\kern-.05em \resizebox{!}{\dimen0}{\raisebox{\depth}{++}}}}
\newcommand{\CCC}{%
    {\settoheight{\dimen0}{C}C/C\kern-.05em \resizebox{!}{\dimen0}{\raisebox{\depth}{++}}}}
\newcommand{\CS}{%
    {\settoheight{\dimen0}{C}C\kern-.05em \resizebox{!}{\dimen0}{\raisebox{\depth}{\#}}}}

\graphicspath{{Figs/}}

\journal{osajournal}

\articletype{Research Article}

\begin{document}
    \suppressfloats 
    
    \title{Benchmarking the Gerchberg-Saxton Algorithm}
    
    \author{Peter J. Christopher,\authormark{1,*} George S. D. Gordon,\authormark{2} and Timothy D. Wilkinson\authormark{1}}
    
    \address{\authormark{1}Centre of Advanced Photonics and Electronics, University of Cambridge\\
        \authormark{2}Department of Electrical and Electronic Engineering, University of Nottingham}
    
    \email{\authormark{*}pjc209@cam.ac.uk}
    
    \homepage{http:\textbackslash\textbackslash www.peterjchristopher.me.uk}
    
    \begin{abstract*}
        Due to the proliferation of spatial light modulators, digital holography is finding wide-spread use in fields from augmented reality to medical imaging to additive manufacturing to lithography to optical tweezing to telecommunications.  There are numerous types of SLM available with a multitude of algorithms for generating holograms.  Each algorithm has limitations in terms of convergence speed, power efficiency, accuracy and data storage requirement.
        
        Here, we consider probably the most common algorithm for computer generated holography - Gerchberg-Saxton - and examine the trade-off in convergent quality, performance and efficiency. In particular, we focus on measuring and understanding the factors that control runtime and convergence. 
    \end{abstract*}

    \section{Introduction}
    
    The primary focus of this paper is on benchmarking the Gerchberg-Saxton (GS) algorithm, perhaps the most common computer generated holography (CGH) algorithm for projection applications using phase modulating devices. 
    
    While many high performance algorithms are available\cite{SortedPixelSelection,STTM,HPS1,HPS2,SSQ}, less information is available on the precise factors influencing performance. In order to support our ongoing research into high power laser projectors this paper benchmarks the widely used GS algorithm. GS  is widely used in relatively "smooth" problems such as display on multi-level SLMs \cite{gerchberg1972practical}. As we shall show later, GS fails to converge for binary SLMs. 
    
    This work begins by discussing the test setup and constraints and discusses individual algorithm components including quantisation, error metrics, starting points and floating point precision. We then continue to expand in detail on performance factors for GS and develop a heuristic relationship for hologram generation performance. 
    
    \section{Considerations}
        
    \subsection{Hardware}
    
    While a number of hardware architectures are available for hologram generation \cite{HardwareReview}, Graphics Processing Units~(GPUs) are the most used and most flexible. Originally targeted for video games, they are seeing increasing use in the scientific and financial sectors. Other works have discussed the relative benefits of alternative architectures such as Field Programmable Gate Arrays~(FPGAs) or Digital Signal Processing Units~(DSPs) in great depth. This work seeks, instead, merely to compare the algorithms used across the different device types. While some quantitative values for generation time will be given, this paper is primarily focussed on relative performance of algorithms on GPU devices.
    
    All the tests discussed here were run on a GTX 1080 GPU with additional system details presented in Table~\ref{tab:hardware}. The processes were run on an independent machine with no additional workload and each test was run many times to ensure consistency and provide estimates of variance. By using low-level interfaces through \CCC, it was possible to manage potential memory bottlenecks to ensure fairer testing. 
    
    The GPU was accessed directly through the native CUDA interface to reduce potential issues with third party libraries. The only major library used is the CUDA Fast Fourier Transform Library~(cuFFT) which is developed by the GPU manufacturer, Nvidia. Independent performance tests are discussed below.
    
    \begin{table}[tbhp]
        \centering
        \caption{Benchmarking hardware}
        \label{tab:hardware}
        \begin{tabular}{cl}
            \toprule
            Workstation & Details                                                                                                                             \\
            \midrule
            GPU         & GTX 1080, 2560 cores, 8GB DDR5 ram, 1607MHz clock                                                                                   \\
            CPU         & Intel\textsuperscript{\textregistered} i7-7700K, overclocked to 4.5GHz, 4 cores, hyperthreaded to 8 cores                           \\
            RAM         & 48GB 2400MHz DDR4                                                                                                                   \\
            OS          & Windows 10 Pro, Build 10.0.17134                                                                                                    \\
            HD          & 500GB M.2 SSD                                                                                                                       \\
            \bottomrule
        \end{tabular}
    \end{table}
    
    \subsection {Test Images}
    
\begin{figure}[tbhp]
        \centering
        \captionsetup{justification=centering}
        \begin{subfigure}[t]{0.31\textwidth}
            \includegraphics[width=\textwidth]{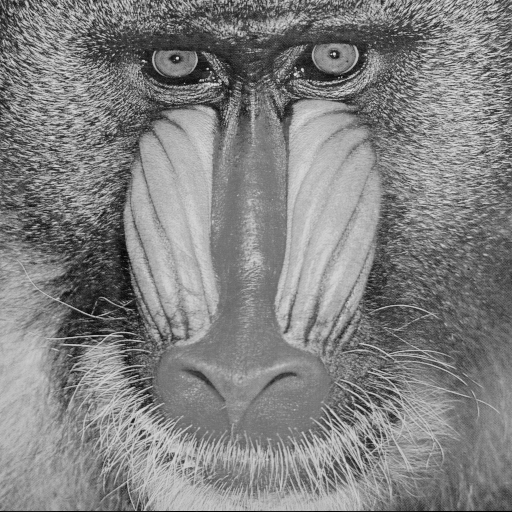}
            \label{fig:TestImage2}
        \end{subfigure}
        \hspace{0.01\textwidth}
        \begin{subfigure}[t]{0.31\textwidth}
            \includegraphics[width=\textwidth]{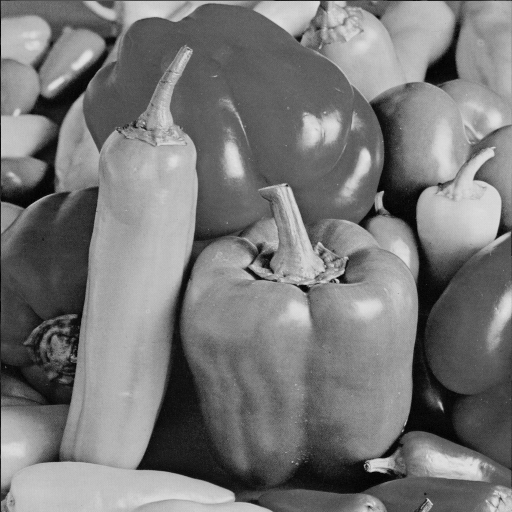}
            \label{fig:TestImage3}
        \end{subfigure}
        \hspace{0.01\textwidth}
        \begin{subfigure}[t]{0.31\textwidth}
            \includegraphics[width=\textwidth]{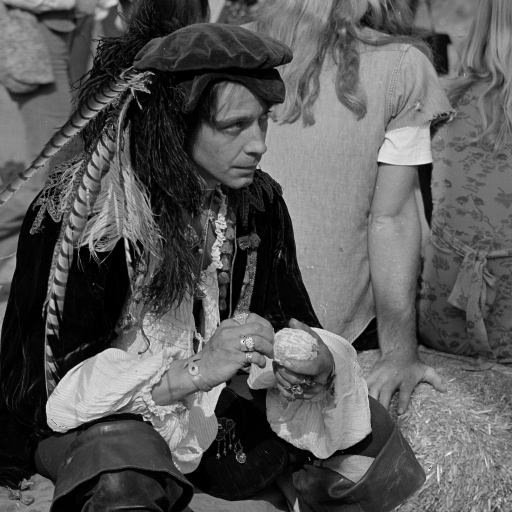}
            \label{fig:TestImage4}
        \end{subfigure}
        \begin{subfigure}[t]{0.31\textwidth}
            \includegraphics[width=\textwidth]{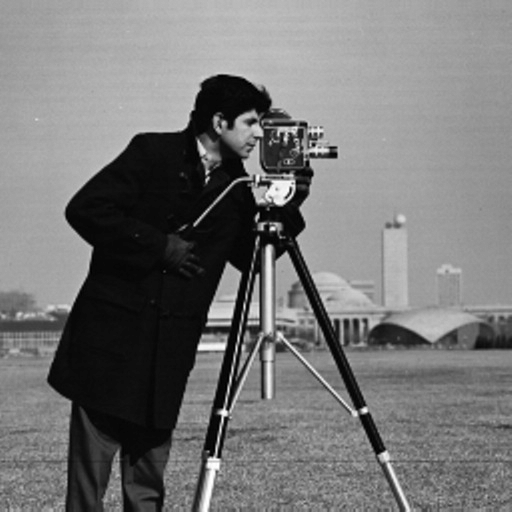}
            \label{fig:TestImage5}
        \end{subfigure}
        \hspace{0.01\textwidth}
        \begin{subfigure}[t]{0.31\textwidth}
            \includegraphics[width=\textwidth]{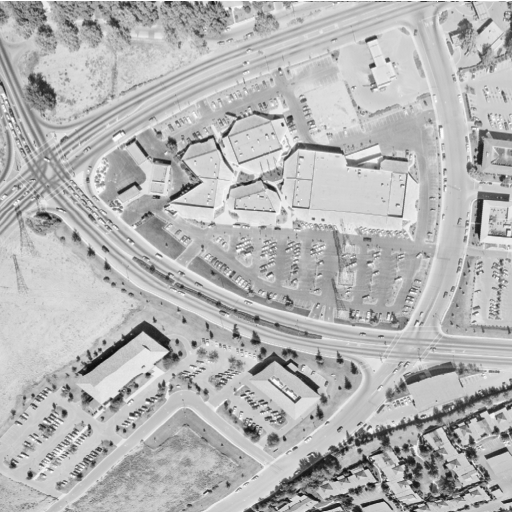}
            \label{fig:TestImage6}
        \end{subfigure}
        \hspace{0.01\textwidth}
        \begin{subfigure}[t]{0.31\textwidth}
            \includegraphics[width=\textwidth]{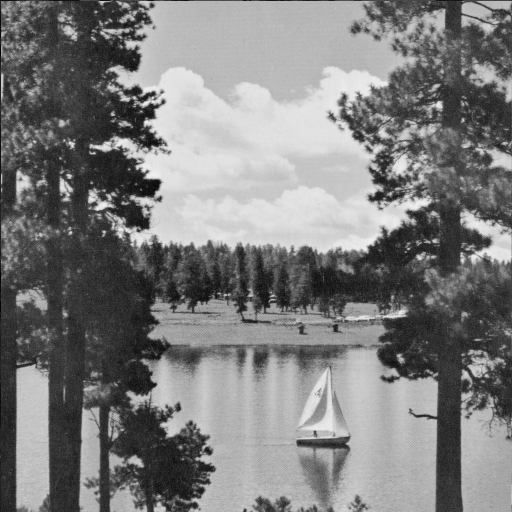}
            \label{fig:TestImage7}
        \end{subfigure}
        \caption{Standard test images from the USC-SIPI image database.\cite{sipidatabaseref} From left to right: Mandrill, Peppers, Man, Camera Man, Aerial, Landscape}
        \label{fig:TestImage}
    \end{figure}

    For the tests carried out we selected the six images from the USC-SIPI image database.~\cite{sipidatabaseref} shown in Figure~\ref{fig:TestImage}. The test images used are 512x512 pixels and, unless otherwise stated, the performance metrics are given for this size. We discuss the effect of image resolution below.
    
    \subsection{FFT Performance}
    
    The algorithm implementations discussed here are based on NVidia's cuFFT package.  In an idealised world, FFTs can be generated in time $O(N^2log(N^2))$. This is rarely achievable, however, due to the limitations and ever increasing complexity of real-world devices. While many benchmarks are available for performance, comparing this information between machines is non-trivial.
    
    Traditionally FFT performance is significantly better when the resolution is a power of 2. A test of 500,000 FFTs with random resolutions between $2^11$ and $2^12$ factorised solely into primes larger than 100 concluded that the performance cost is of approximately $2.6-3.1$ times that expected of a resolution power of 2. 
    
    The performance for the 255 unfactorisable primes between $2^11$ and $2^12$ had performances significantly greater than 10 times worse. In order to reduce this factor, every sampling point is equal to a power of 2 multiplied by up to 2 values from the set ${3, 5 and 7}$.
    
\begin{figure}[tbhp]
        {\includegraphics[trim={0 0 0 0},width=1.0\linewidth,page=1]{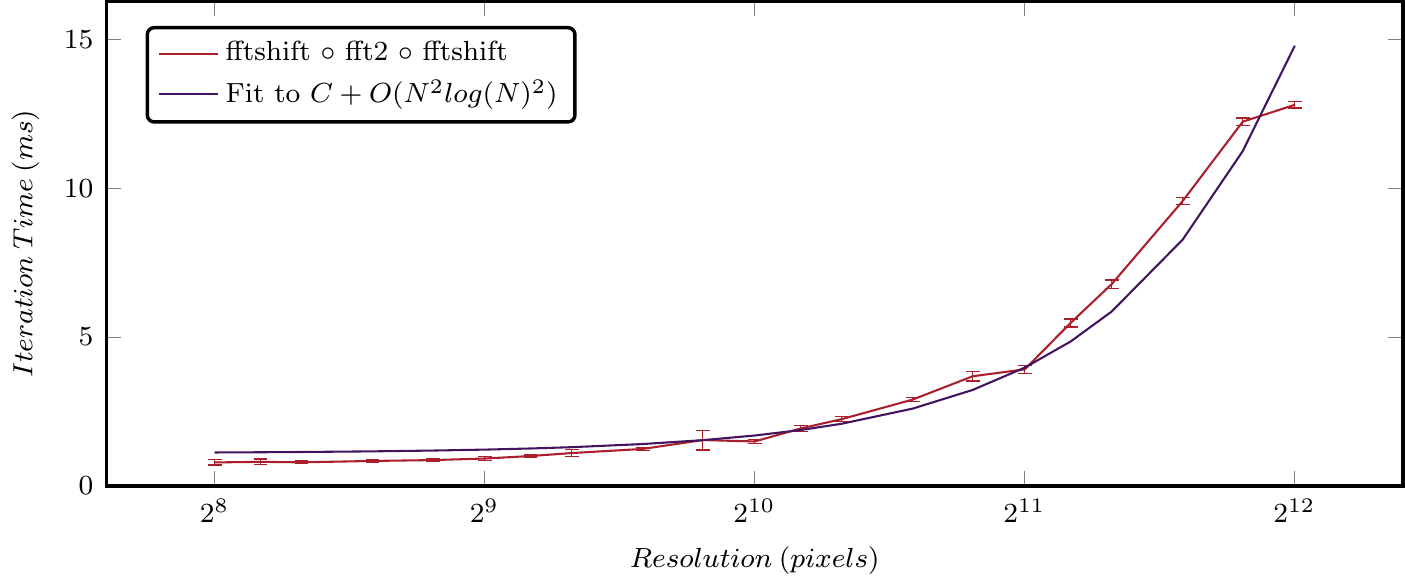}}
        \caption[FFT Performance]{Comparison of FFT performance against image resolution using Nvidia cuFFT. Error bars show the $2\sigma$ confidence interval measured from 100 independent runs of 1000 pairs of FFTs and IFFTs}
        \label{graph:fft1}
    \end{figure}

    Figure~\ref{graph:fft1} shows the behaviour of cuFFT against resolution along with the $O(N^2log(N^2))$ trend line. The drop in the final measurement appears to be due to cuFFT making the transition to the 64 bit kernel.
    
    The FFT calculations took up $>98\%$ of the algorithm runtime of all the algorithms discussed here. Of the $>100,000,000$ FFTs run during testing, $<2\%$ of the runtime was spent on memory movements onto the GPU, transpose operations and quantisations. For analysis, therefore, the FFT time is assumed to dominate. Additionally, the shift, transpose and quantisation steps all na\"ively run in $O(N^2)$ time.
    
    A two dimensional FFT looks like the following:
    
    \begin{equation} \label{fftmult1}
    \texttt{fft2} \rightarrow \texttt{transpose} \circ \texttt{fft} \circ \texttt{transpose} \circ \texttt{fft}
    \end{equation}
    
    FFTs as used in holography require the zero order to be shifted to the centre of the image. This requires two additional $\texttt{fftshift}$ operations
    
    \begin{equation} \label{fftmult2}
    \texttt{fftshift} \circ \texttt{transpose} \circ \texttt{fft} \circ \texttt{transpose} \circ \texttt{fft} \circ \texttt{fftshift}
    \end{equation}
    
    where $\texttt{fftshift}$ is an operation to swap diagonal quadrants of an image.
        
    \subsection{Floating Point Precision}
    
    Traditional CPU arithmetical calculations are done using a math coprocessor or Floating Point Unit~(FPU). On modern Intel x86/x86-64 systems, the precision of the floating points used make little difference to calculation time and double (64-bit) precision is standard. On Nvidia graphics cards, single precision (32-bit) offers performance boosts of $~ 2 \times $ over double precision. Modern GPUs~(SM\_53 or later) are capable of working in half precision though the speed increase is $<2\times$. 
    
    HoloGen is capable of being compiled in double, single and half precision variants. Running a suite of tests on the workstation machine described in Table~\ref{tab:hardware} on all three precisions showed that single precision was $\approx 100\%$ faster than double precision in almost all cases while half precision only offered $\approx 30-50\%$ speed improvements. This is summarised in Table~\ref{tab:precision}. 
    
    The half precision performance in expected to improve with GPU generation and offers significant promise for hologram algorithms. The linearity of the FT means that values can be normalised near to 1, reducing the impact of reduced exponent bits. Incremental algorithms such as GS only run for $10-100$ cycles, reducing the impact of accumulated errors while longer algorithms like DS which can run to $>100,000$ cycles don't introduce incremental errors.
    
    The results shown here are generated for single precision as this is the standard in similar analysis.
    
    \begin{table}[tbhp]
        \centering
        \caption[Floating Point Precision]{Impact of floating point precision on hologram generation speeds}
        \label{tab:precision}
        \begin{tabular}{ccl}
            \toprule
            & Time      &                                                   \\
            Precision       & Modifier  & Description                                       \\
            \midrule
            Single (32-bit) & 1         & 1 sign bit, 8 exponent bits, 24 significand bits  \\
            Double(64-bit)  & 1.96-1.99 & 1 sign bit, 11 exponent bits, 53 significand bits \\
            Half (16-bit)   & 0.67-0.76 & 1 sign bit, 5 exponent bits, 11 significand bits  \\
            \bottomrule
        \end{tabular}
    \end{table}

    \subsection{Quantisation} 
    
    Spatial Light Modulators~(SLMs) are unable to arbitrarily modulate light, instead typically modulating only in discrete values of phase or amplitude. The act of modifying a target light field to meet these constraints is here referred to as \textit{quantisation}.    
    
    By far the most dominant quantisation approach is the \textit{nearest-neighbour} approach where the pixel is quantised to the nearest complex value achievable on the device. Other schemes such as \textit{error diffusion} exist but can be considered algorithms in their own right and are beyond the scope of this paper. As modern SLMs are usually discretely addressed, they present a number of modulation levels which typically vary between $2^1$ and $2^8$. 
    
    \subsection{Test Setup}
    
\begin{figure}[tb]
        \centering
        \captionsetup{justification=centering}
        \begin{subfigure}[t]{0.18\textwidth}
            \includegraphics[width=\textwidth]{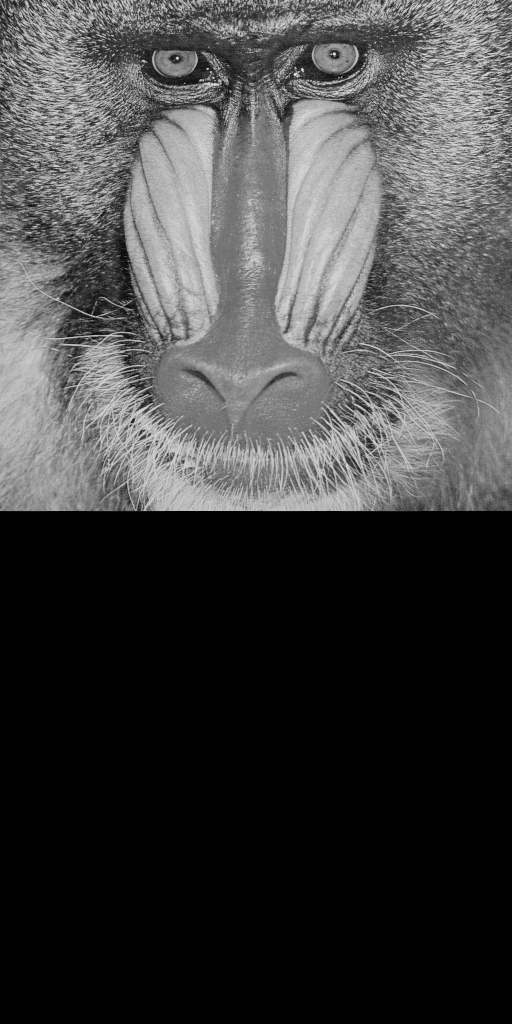}
            \caption{Target}
        \end{subfigure}
        \hspace{0.01\textwidth}
        \begin{subfigure}[t]{0.18\textwidth}
            \includegraphics[width=\textwidth]{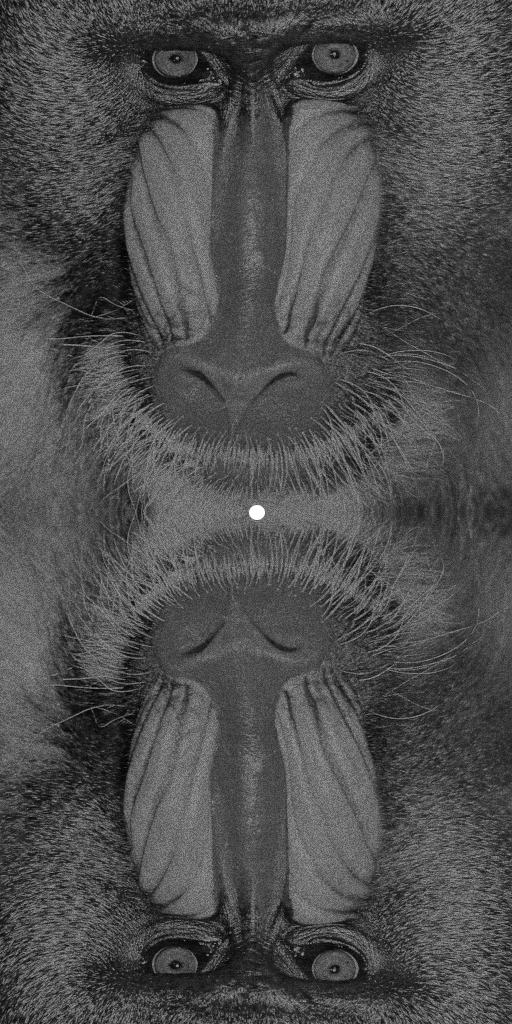}
            \caption{Multi-level Amplitude}
        \end{subfigure}
        \hspace{0.01\textwidth}
        \begin{subfigure}[t]{0.18\textwidth}
            \includegraphics[width=\textwidth]{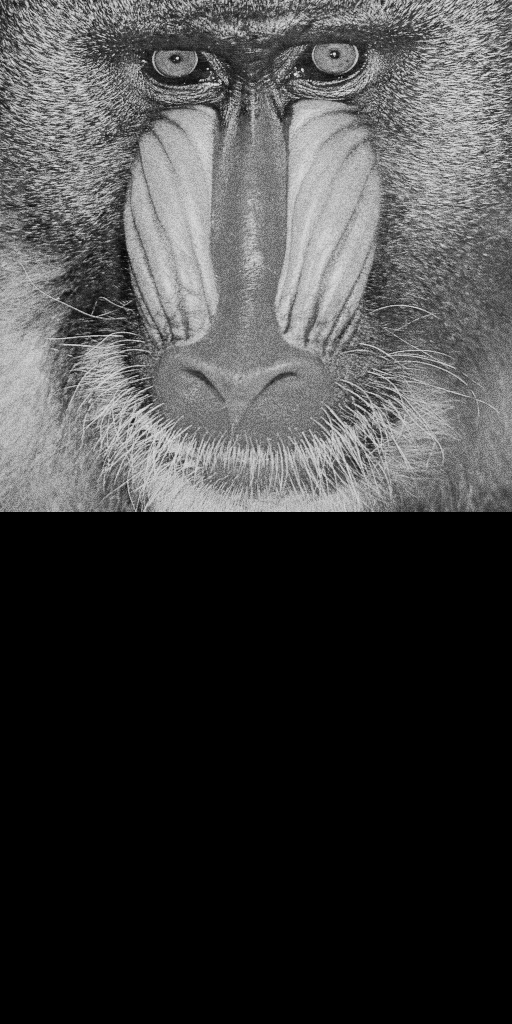}
            \caption{Multi-level Phase}
        \end{subfigure}
        \hspace{0.01\textwidth}
        \begin{subfigure}[t]{0.18\textwidth}
            \includegraphics[width=\textwidth]{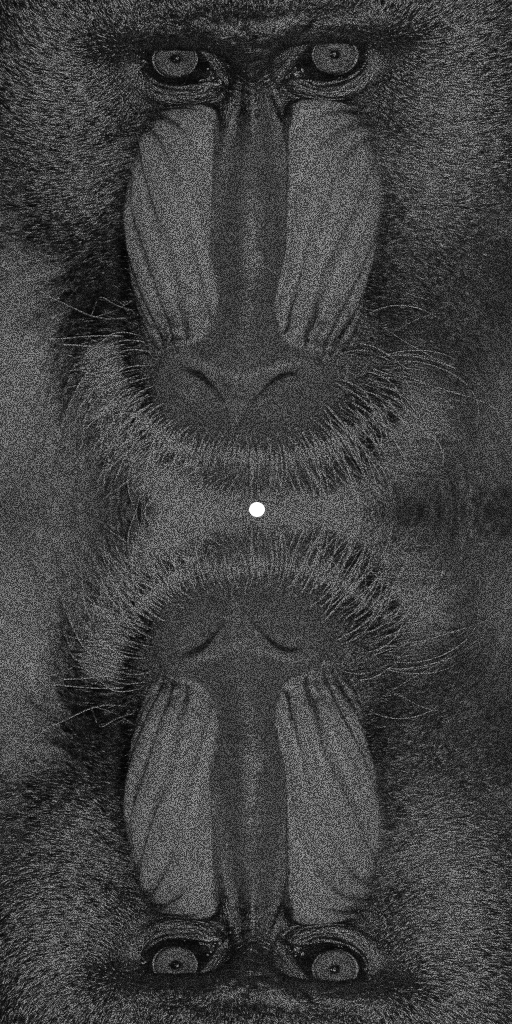}
            \caption{Binary Amplitude}
        \end{subfigure}
        \hspace{0.01\textwidth}
        \begin{subfigure}[t]{0.18\textwidth}
            \includegraphics[width=\textwidth]{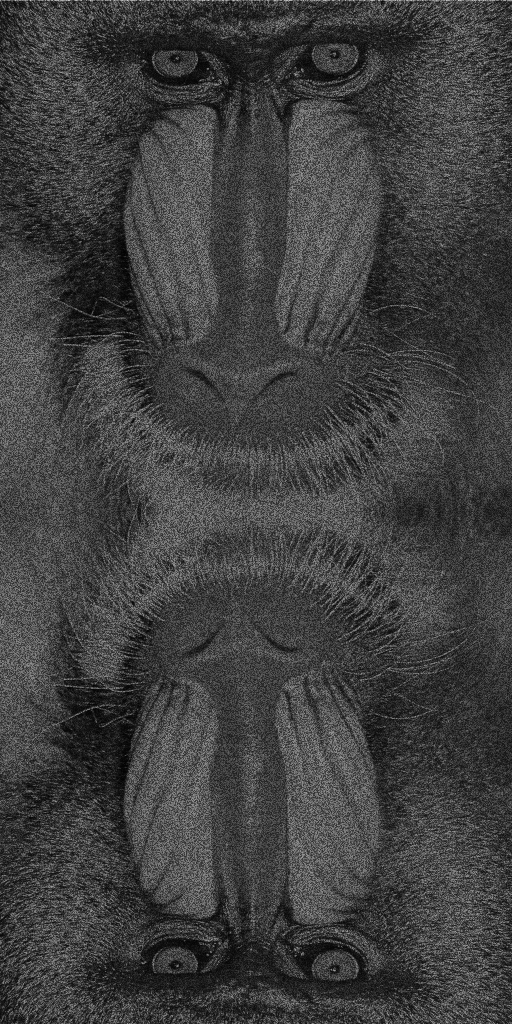}
            \caption{Binary Phase}
        \end{subfigure}
        \caption{Benchmarking test setup showing the target image (a) and the symmetry and zero order effects of different modulation schemes (b-e).}
        \label{fig:format}
    \end{figure}

    The majority of simulations run assumed an SLM twice the size of the target image. This is to compensate for the rotational symmetry inherent in binary holograms as shown in Figure~\ref{fig:format}. This corresponds to the format shown in Figure~\ref{fig:format}~(a). Different SLM modulation schemes offer different challenges in terms of conjugate symmetry and zero order behaviour. 
    
    \subsection{Starting Point}
    
    For the algorithms studied later, the starting point is often a significant factor in convergence time. Two main approaches are often used:
    
    \begin{itemize}
        \item \textbf{Randomisation} - Random points in the unit circle are generated for each diffraction field pixel. These are then quantised to the SLM modulation behaviour.
        \item \textbf{Back Projection} - The target replay field is back projected to the diffraction field and quantised to the SLM modulation behaviour.
    \end{itemize}
    
    For many algorithms, back projection is preferable for a single quick solution with randomisation being used for better quality, slower results. While other starting points exist, they are little used and can be considered beyond our scope.
    
    \subsection{Merit Function and Error}
    
    During hologram generation, tuning is achieved by adjusting the \textit{three freedoms}: \textit{amplitude}, \textit{phase} and \textit{scale}. In a typical application, only a portion of the replay field is of interest, giving amplitude freedom in the other regions. Phase freedom is due to the eye being phase insensitive and scale freedom is provided in applications where image fidelity is more important than efficiency. Careful adjustment of these factors can cause several orders of magnitude of difference in processing times~\cite{wyrowski1990diffractive}. It is important to note that it is often mathematically impossible to obtain an exact solution.
    
    The most common approach to measuring hologram performance is that of Mean Squared Error~(MSE). The MSE $E_{MSE}(T,R)$ is given as a relation of the target image $T$ and generated replay field $R$. 
        
    \begin{equation} \label{mse}
        E_{\text{MSE}, \text{PI}}(T,R) = \frac{1}{N_x N_y}\sum_{x=0}^{x=N_x-1}\sum_{y=0}^{y=N_y-1} \left[\abs{T(x,y)} -  \abs{R(x,y)}\right]^2 
    \end{equation}
     
    This relationship assumes that the solution is \textit{phase insensitive} and that only the spatial intensity profile is of interest. While this is true for display applications, many applications also require phase control and can use the following formulation of MSE.
    
    \begin{equation} \label{mse2}
        E_{\text{MSE}, \text{PS}}(T,R) = \frac{1}{N_x N_y}\sum_{x=0}^{x=N_x-1}\sum_{y=0}^{y=N_y-1} \abs{T(x,y) -  R(x,y)}^2 
    \end{equation}
    
    For displays viewed by the human eye, the Structural Similarity Index~(SSIM) is also used as it has been shown to more closely correspond to ocular visual quality.\cite{wang2004image}
        
    \begin{equation}
        E_{\text{SSIM}}(T,R) = \frac{\left(2\mu_T\mu_R+c_1\right)\left(2\sigma_{TR}+c_2\right)} {\left(\mu_T^2+\mu_R^2+c_1\right)\left(\sigma_T^2+\sigma_R^2+c_2\right)}
    \end{equation}
        
    where $\mu_T$ and $\mu_R$ are the target and replay means; $\sigma_T$ and $\sigma_R$ are the target and replay variances; $\sigma_{TR}$ is the covariance of the two images and $c_1$ and $c_2$ are functions of pixel dynamic range, $L$, where $c_1=(k_1L)^2$ and $c_2=(k_2L)^2$. $k_1$ and $k_2$ are usually taken as $0.01$ and $0.03$ respectively.
        
    MSE is a poor metric of image quality to a human observer of noise in a holographic image but is often used when assessing algorithm performance.~\cite{buckley2011real, buckley200870}
        
    In this work we exclusively use error as PI MSE $E = E_{MSE, PI}$ but it should be noted that other formulations are available. Additionally, Normalised MSE~(NMSE) is introduced where the MSE of each test image is normalised to that of the benchmark \textit{Mandrill} test image by the ratio of the convergent graphs. This allows for using a variety of test images while keeping the results simple to understand.
    
    \begin{align}
    NMSE_{image,n} = MSE_{image,n} \frac{MSE_{Mandrill,1000}}{MSE_{image,1000}} 
    \end{align}
    
    \subsection{Convergence}
    
    Mathematically a series $x_{n}$ is \textit{convergent} to value $L$ if for any given $\epsilon$ there exists a value $n$ for which the following applies
    
    \begin{equation}
    \abs{x_t - L} < \epsilon \quad \forall \quad t \ge n
    \end{equation}
    
    and \textit{diverges} otherwise. This is not practical for use in a non-linear piecewise context.
    
    Instead, for this application, the series NMSE is taken as converging to value $L$ being the first series element $n$ for which the following applies. 
    
    \begin{equation}\label{eqn:convergence}
    \abs{NMSE - L} < \epsilon \quad \forall \quad 2n \ge t \ge n
    \end{equation}
    
    The value of $\epsilon$ is defined arbitrarily to be $0.001$. 
    
    \subsection{Data and Code Availability}
    
    The results of all the tests run are available in a JavaScript Object Notation~(JSON) files available in the online supplementary material. An in-house application, HoloGen, is used for running the tests. HoloGen is an early-access open-source application available from \url{https://gitlab.com/CMMPEOpenAccess/HoloGen}. HoloGen is licensed under the MIT license \cite{model12s2} and the tests here were run on build v2.2.1.17177.
    
    \section{Gerchberg-Saxton Algorithm and Variants}
    
    This section discusses the first algorithm under consideration: Gerchberg-Saxton. Originally presented in 1972 as a means of phase retrieval,~\cite{gerchberg1972practical, hirsch1971method} Gerchberg-Saxton~(GS) quickly became a common means of hologram generation.~\cite{wyrowski1988iterative, fienup1986phase} 
    
\begin{figure}[tbhp]
        {\includegraphics[trim={0 0 0 0},width=1.0\linewidth,page=1]{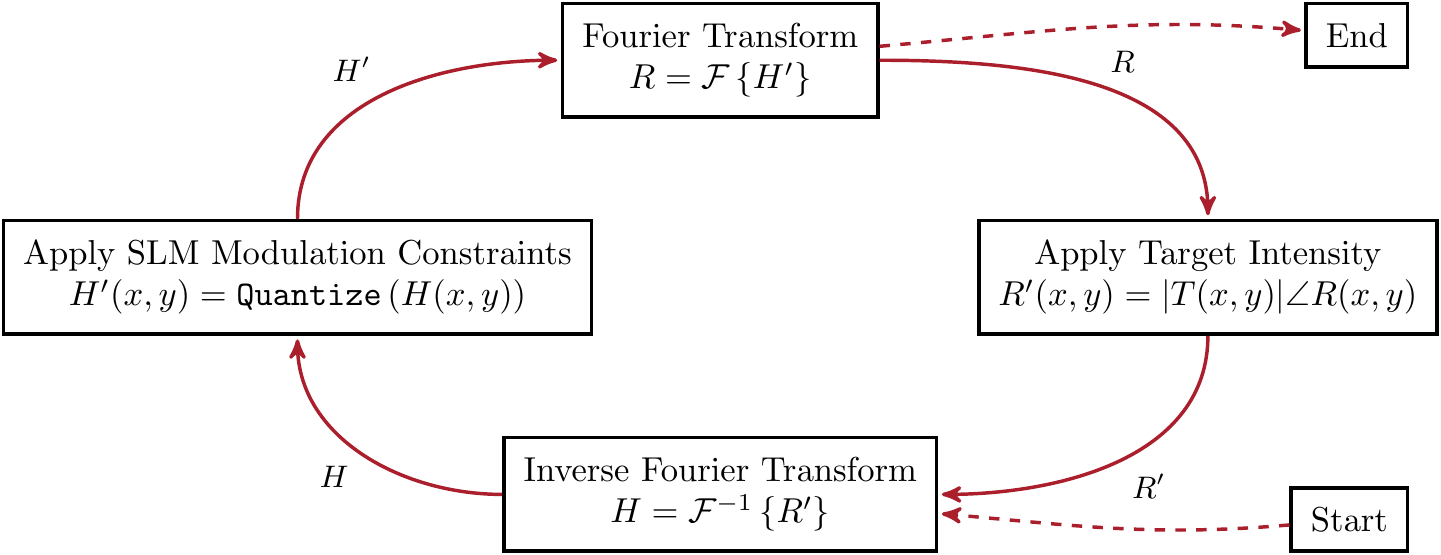}}
        \caption{Gerchberg-Saxton Algorithm}
        \label{fig:alg-1}
    \end{figure}

    GS is part of the wider family of Iterative Fourier Transform Algorithms~(IFTAs) which operate by transforming an image between the diffraction field and replay field while enforcing the constraints of both as shown in Figure~\ref{fig:alg-1}. It can be shown that the IFTA approach minimises MSE as well as producing identically independently distributed~(i.i.d.) results for i.i.d. randomised phase inputs.~\cite{sinclair2004interactive, sypek2005three, chang2014holographic} This only applies, however, in the unmodulated case. For modulated cases divergence is common.
    
\begin{figure}[tbhp]
        {\includegraphics[trim={0 0 0 0},width=1.0\linewidth,page=1]{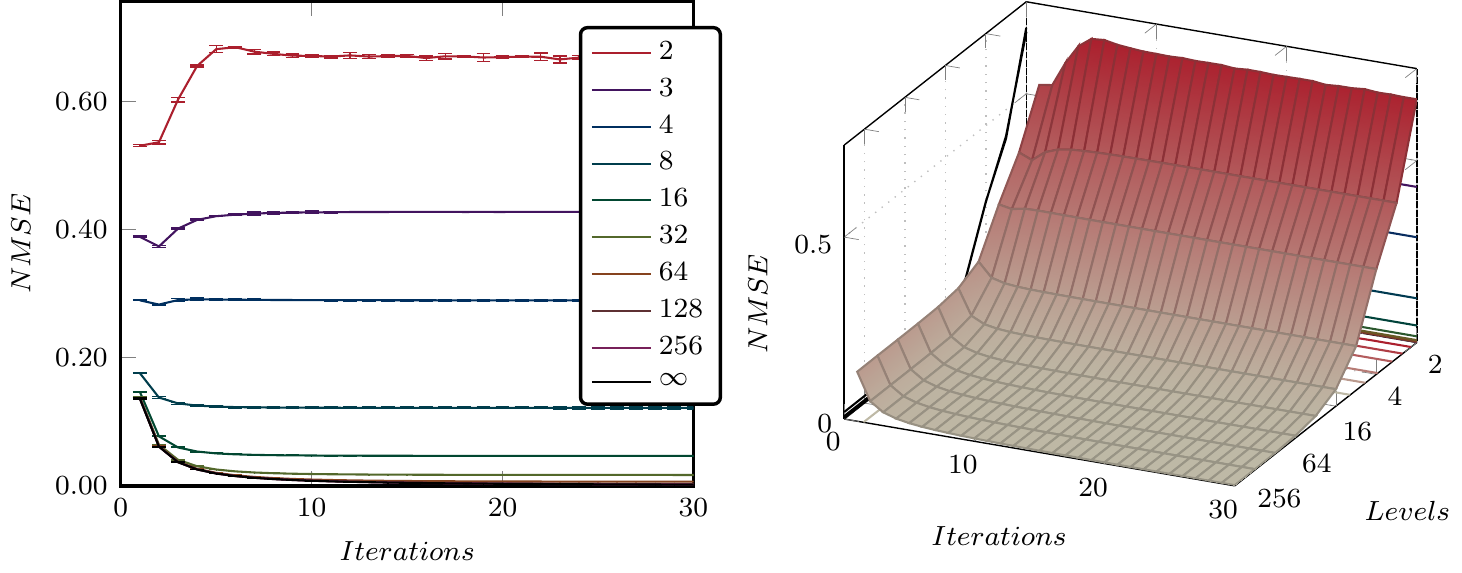}}
        \caption{Gerchberg-Saxton convergence with randomised start for numbers of modulation levels. Error bars show the $2\sigma$ confidence interval. Each line is taken as the mean of 50 independent runs for each of the 6 standard test images shown in Figure~\ref{fig:TestImage}. }
        \label{GSErrorVsItrPerLevelPhaseRandom}
    \end{figure}

\begin{figure}[tbhp]
        {\includegraphics[trim={0 0 0 0},width=1.0\linewidth,page=1]{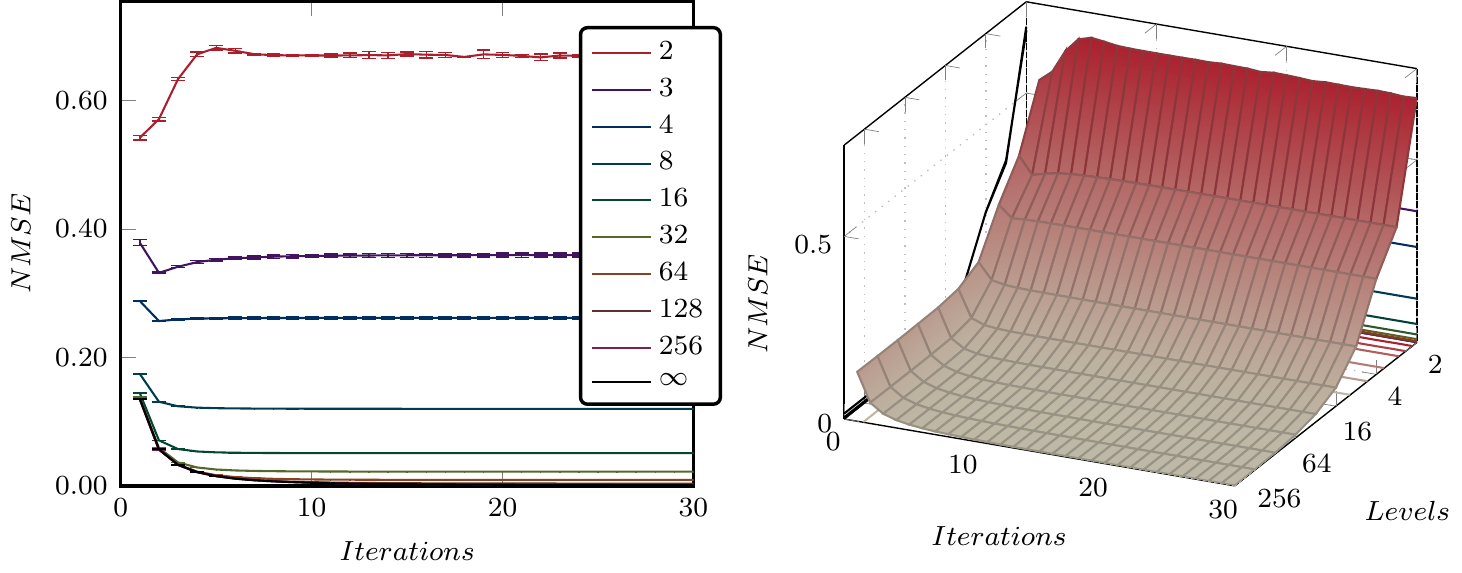}}
        \caption{Gerchberg-Saxton convergence with back projected start for numbers of modulation levels. Error bars show the $2\sigma$ confidence interval. Each line is taken as the mean of 50 independent runs for each of the 6 standard test images shown in Figure~\ref{fig:TestImage}. }
        \label{GSErrorVsItrPerLevelPhaseBackProject}
    \end{figure}

    Figure~\ref{GSErrorVsItrPerLevelPhaseRandom} shows the convergence with $2\sigma$ confidence interval for the test images shown in Figure~\ref{fig:TestImage}. Each line is taken as the average of 50 runs for each of the 6 standard images and 10 modulation levels over 30 iterations. A total of 3,000 independent runs. The simulated device is phase modulating and a randomised starting point is used. This can be compared with Figure~\ref{GSErrorVsItrPerLevelPhaseBackProject} where an IFFT of the target image is used, a so called back projected starting point. It will be seen that convergence for GS is extremely fast, plateauing after only about $10$ iterations. 
    
    For binary holograms, and other low level count holograms, cycles were often observed in the solution. Rather than converging to a single static result, a cycle of upwards of 10 elements would form. This periodic solution could have significant impact on the MSE with variations of up to 9\%. For optimum results with GS, it should not be assumed that MSE always improves.
    
\begin{figure}[tbhp]
        {\includegraphics[trim={0 0 0 0},width=1.0\linewidth,page=1]{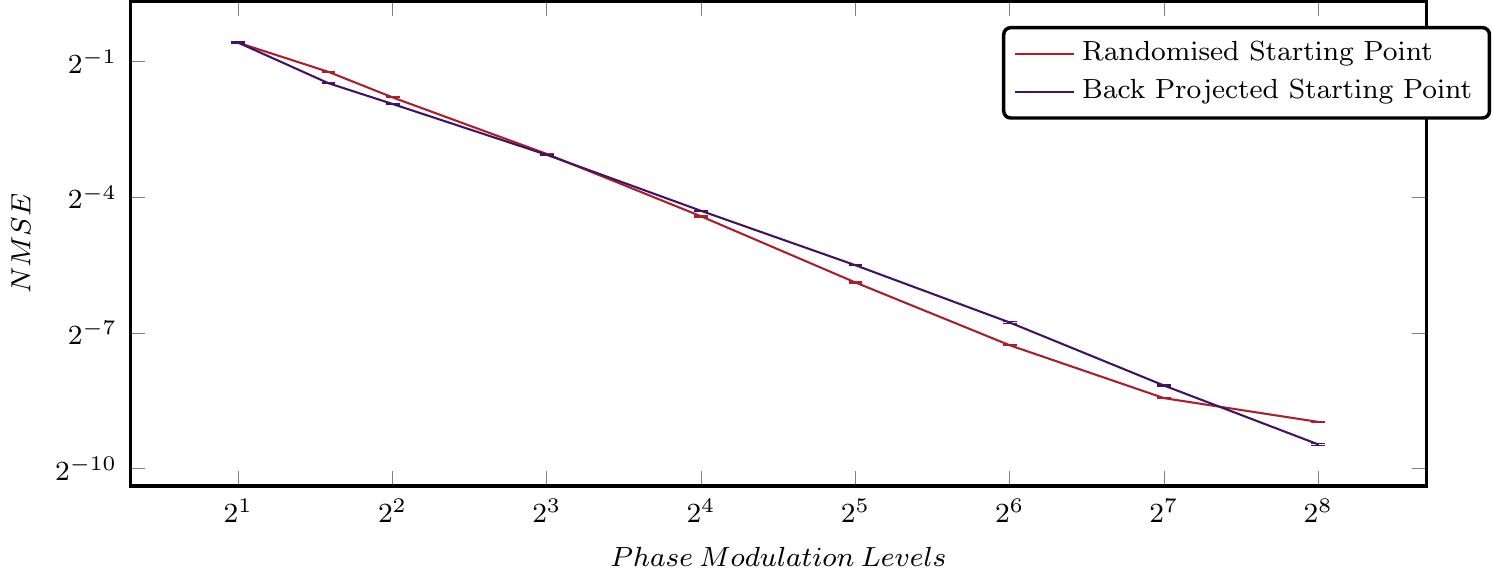}}
        \caption{Gerchberg-Saxton convergent error against number of modulation levels for starting points. Error bars show the $2\sigma$ confidence interval. Each line is taken as the mean of 50 independent runs for each of the 6 standard test images shown in Figure~\ref{fig:TestImage}. }
        \label{GSPhaseErrorVsLevelCrossSection}
    \end{figure}

    Another observation confirms that by Cable \cite{Cable06} that beyond a certain point , more modulation levels does not improve the quality of the hologram. A continuously modulated hologram offers less than a 1\% improvement on a 64 level hologram. This can be seen in Figure~\ref{GSPhaseErrorVsLevelCrossSection}. Certainly, there is no need for greater than 8-bit addressed real-world systems (256 levels).

    A log-log plot of convergent error (where convergence is taken as being given according to Eq.~\ref{eqn:convergence}) against number of modulation levels is shown in Figure~\ref{GSPhaseErrorVsLevelCrossSection}. The linear relationship is of interest and the authors are unaware of it having been reported previously. This result is not unexpected for the system, but it is good to see it confirmed here.

\begin{figure}[tbhp]
        {\includegraphics[trim={0 0 0 0},width=1.0\linewidth,page=1]{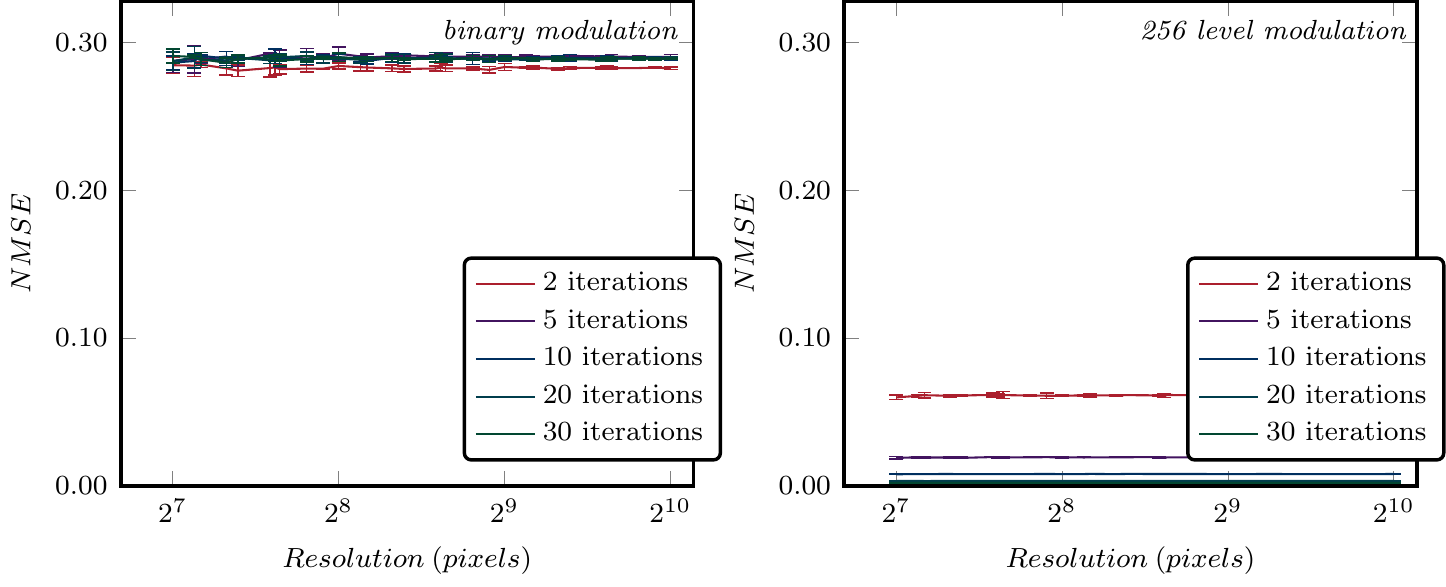}}
        \caption{Gerchberg-Saxton error against resolution for binary (left) and 256 (right) level modulation. Error bars show the $2\sigma$ confidence interval. Each line is taken as the mean of 50 independent runs for each of the 6 standard test images shown in Figure~\ref{fig:TestImage}. }
        \label{GSPhaseErrorVsResPerItr}
    \end{figure}

    The behaviour against resolution is also of interest. Figure~\ref{GSPhaseErrorVsResPerItr} shows the error for a binary~(left) and multi-level~(right) device after set numbers of iterations for resolutions ranging between $128\times128$ up to $1024\times1024$. The flatness of the lines is of interest, suggesting the final pixel error and number of iterations until convergence are both independent of the resolution of the test image.

\begin{figure}[tbhp]
        {\includegraphics[trim={0 0 0 0},width=1.0\linewidth,page=1]{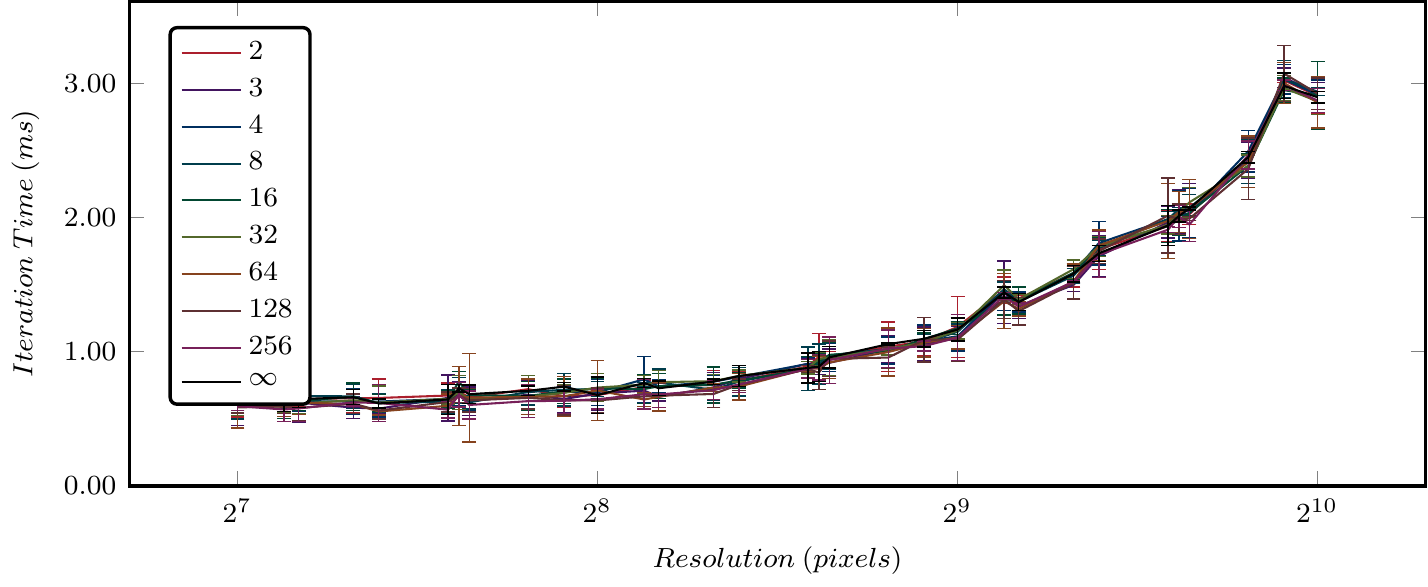}}
        \caption{Gerchberg-Saxton iteration time against resolution for numbers of modulation levels. Error bars show the $2\sigma$ confidence interval. Each line is taken as the mean of 50 independent runs for each of the 6 standard test images shown in Figure~\ref{fig:TestImage}. }
        \label{GSPhaseTimeVsResPerLevel}
    \end{figure}

    Finally, Figure~\ref{GSPhaseTimeVsResPerLevel} shows the iteration time against resolution. The jump in the graph at $N_x=N_y\approxeq 2^{8.5} \approxeq 360$ is due to a similar jump in the calculation time in the cuFFT library. Excluding this jump, as expected for a cuFFT based process, the graph fits with $<5\%$ MSE to the expected $O\left(N^2 \log(N)\right))$ curve.
    
    \subsection{Performance} \label{gsperf}
    
    While an analytical relationship of problem performance is impossible, an approximate formula can be used for estimating process performance for given a system and parameters. Using the $>100$ million iterations calculated, we suggest the following approximate formula that may be used for estimating run time $t$ for a given system. 
    
    \begin{equation}
    t_{GS} = C_{numitr} C_{machine} C_{precision} C_{software}(C_{itr_1}  + C_{itr_2} N_x N_y \: Log(N_x)Log(N_y)) \si{\milli\second}
    \end{equation}
    
    where 
    
    \begin{itemize}
        \item $C_{itr_1} \approx 0.71$ and represents the memory transfer and input/output component of the operation. 
        \item $C_{itr_2} \approx 1.09 \times 10^{-6}$ and represents a scaling constant  
        \item $C_{machine}$ is a parameter of the machine used and is equal to $1$ in this case. As the process is GPU bound, $C_{machine}$ can be treated as $C_{machine} \propto \frac{GFLOPS}{102.8}$ where $102.8 \times 10^9$ is the base FLOP rate for the GTX 1080 test device.
        \item $C_{software}$ is a parameter of the code developed and is equal to $1$ in this case. As the software is applied to the general case and, it is estimated that this could be straightforwardly reduced to $0.5$ for very specific cases.
        \item $C_{numitr}$ is the number of iterations.  
        \item $C_{precision}$ is a factor for floating point precision and is equal to $1$ in this case. $C_{precision}$ can be determined from Table~\ref{tab:precision}.
    \end{itemize} 
    
    While all of these parameters can be expected to vary from system to system, this heuristic relationship can be used as a starting point for system design. Verifying this relationship for all devices is impossible but tests were run on seven different workstations varying in purchase date from 2011 to 2019 and with rated GFLOP performances varying by more than two orders of magnitudes. All seven machines demonstrated performances within $20\%$ of the relationship given here with the value of $C_{itr_1}$ being the primary variable.
    
    \subsection{Variants}
    
    Two variants of GS are worthy of notice: Weighted Gerchberg-Saxton and Liu-Taghizadeh.
    
    \begin{figure}[tbhp]
        {\includegraphics[trim={0 0 0 0},width=1.0\linewidth,page=1]{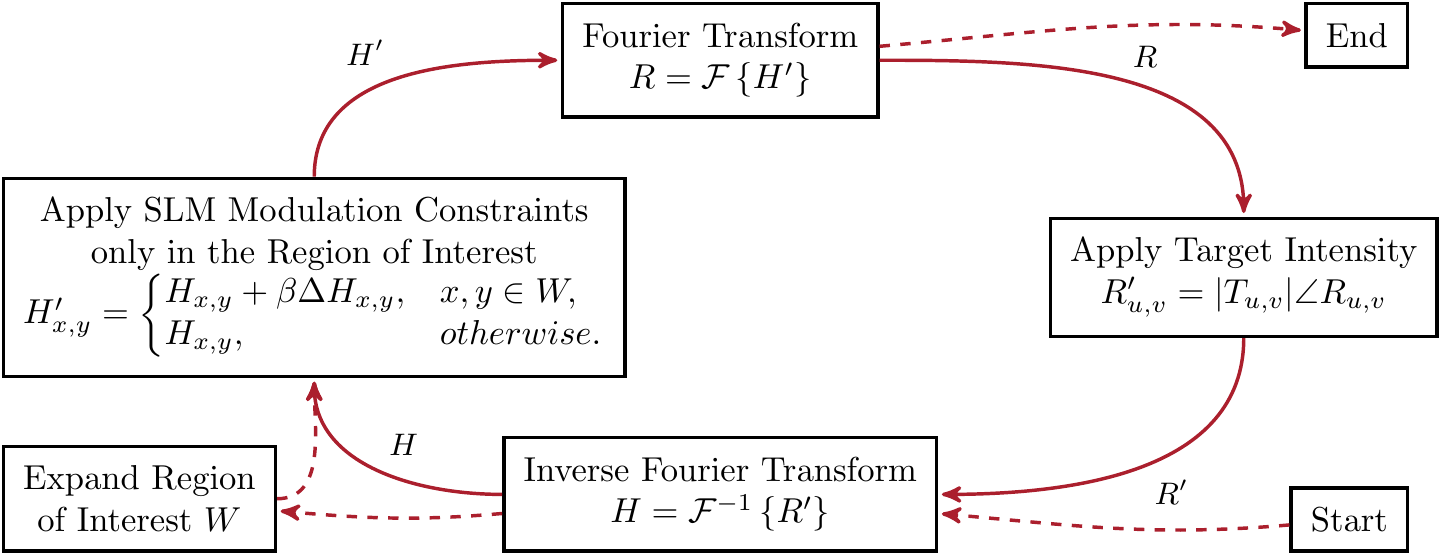}}
        \caption{Weighted Gerchberg-Saxton Algorithm}
        \label{fig:alg-2}
    \end{figure}
    
    \begin{figure}[tbhp]
        {\includegraphics[trim={0 0 0 0},width=1.0\linewidth,page=1]{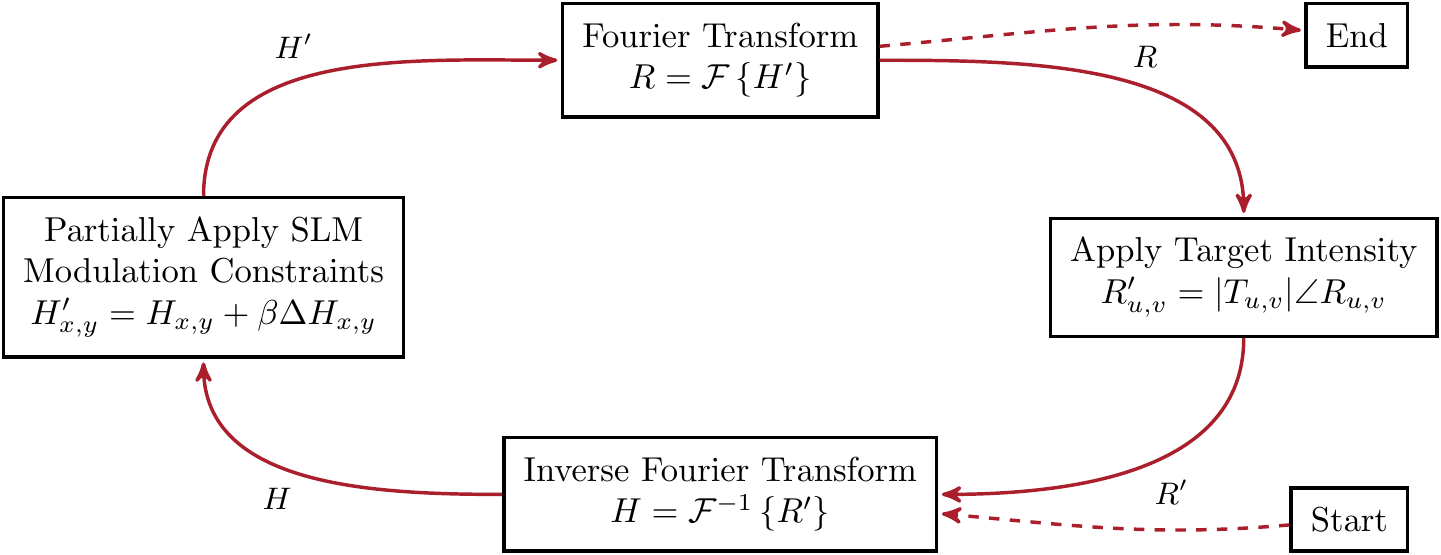}}
        \caption{Liu-Taghizadeh Algorithm}
        \label{fig:alg-3}
    \end{figure}

    Weighted Gerchberg-Saxton operates by partially applying the modulation constraints.~\cite{Johansson2000, Schafer2001, Kuzmenko2011} Here the diffraction side quantisation step where the hologram is constrained to the limits of the SLM is modified to either under- or over-compensate the change by a factor $\beta$.~\cite{wyrowski1989iterative, fienup1982phase} This leads to a relationship
    
    \begin{equation}
    H'  = \texttt{Quantise}\left(H\right)  = H + \beta \Delta H 
    \end{equation}
    
    As with many other GS variants, this can improve convergence speed and convergent quality but adds significant overhead in complexity and user expertise.~\cite{Cheremkhin2014, Biggs1997}  

    The Liu-Taghizadeh~(LT) Algorithm is a well-developed approach that initially restricts the region of interest in the target field
        
    \begin{equation} 
        H'  = \texttt{Quantise}\left(H\right)  = \left\{
        \begin{array}{@{}ll@{}}
        H + \Delta H, & \text{where} \quad (x,y) \in W, \\
        H,                    & \text{otherwise.}               
        \end{array}\right.
    \end{equation}
        
    where $W$ is the region of interest where the constraints are applied and $\Delta H$ is the na\"ive quantization. This reduces the number of degrees of freedom being optimised for at any one point.~\cite{liu2006symmetrical, georgiou2008aspects, liu2002iterative}  LT algorithms require experience and time to use but can offer up to $100\%$ improvements in execution time over native GS in specialised cases.~\cite{Chang2015, liu2002iterative} LT continues to see active research in well-defined problems.~\cite{Qin2009, Tao2015}

    \section{Conclusion}
    
    \textit{Summarise the intro bit}
    
    This paper has benchmarked the Gerchberg-Saxton algorithm performance on a modern workstation. GS was found to offer high performance in the case of multi-level devices but to diverge in the case of binary devices. The number of iterations taken for convergence and the error after a given number of iterations was found not to depend on the target resolution. The starting point for the algorithm was also seen to have little bearing on the convergent error and convergence time. It was also seen that the benefits of increasing the number of modulation levels per pixel was seen to decrease according to a power relationship and it was suggested that the maximum number of modulation levels required by real-world systems is approximately $2^6$.
    
    A heuristic relationship for calculating the expected runtime of the algorithm depending on a wide array of system parameters was also presented. This allows for approximate specification of device requirements as a function of the system, resolution and device floating operations per second~(FLOPs).
        
    
    

    \bibliography{references}

\end{document}